\documentclass[final,5p,times,twocolumn]{elsarticle}

\usepackage{amssymb,amsmath}

\usepackage{url}

\newcommand{\beq}{\begin{equation}}
\newcommand{\eeq}{\end{equation}}

\journal{...}

\begin{document}

\begin{frontmatter}



\title{A quartet BCS-like theory}


\author[1,2,3]{V. V. Baran\corref{cor1}}
 \ead{virgil.baran@theory.nipne.ro} 
\author[1,3,4]{D. S. Delion}
 
 \cortext[cor1]{Corresponding author}
 
 \address[1]{``Horia Hulubei'' National Institute of Physics and 
Nuclear Engineering, 30 Reactorului, RO-077125, Bucharest-M\u agurele, Rom\^ania} 
\address[2]{Faculty of Physics, University of Bucharest,
405 Atomi\c stilor, POB MG-11, RO-077125, Bucharest-M\u agurele, Rom\^ania}
\address[3]{Academy of Romanian Scientists,
54 Splaiul Independen\c tei, RO-050094, Bucharest, Rom\^ania}
\address[4]{Bioterra University, 81 G\^arlei, RO-013724, Bucharest, Rom\^ania}

\begin{abstract}
We introduce a BCS-like theory for the quartet correlations induced by the isovector pairing interaction. It is based on a coherent state of BCS type and, unlike usual mean field approaches, it displays a vanishing pair anomalous density $\langle c^\dagger c^\dagger\rangle =0$. We find good agreement between our theory and the exact results. We discuss how the pairing and quarteting correlations share some similar qualitative features within the BCS approach. However, there is no sharp quarteting phase transition. We also present various ways in which our theory may be further developed.

\end{abstract}


\end{frontmatter}

\emph{Introduction}. The advances in the experimental techniques of the last three decades have opened up new possibilities to investigate the nature of the nuclear interactions at the $N=Z$ line. The significant overlap between neutron and proton orbitals in this region allows the the existence of \emph{proton-neutron (pn) pairing}, which is suppressed away from $N=Z$. 

Already in the first theoretical approach to $pn$ pairing dating six decades ago, Belyaev, Zacharev and Soloviev \cite{Bel60} realized that ``one must take into consideration the quadruple correlation of $\alpha$ particle-like nucleons in addition to pair correlations" \cite{Bel60}. Subsequent works proposed various ways to incorporate these \emph{quartet} correlations into a functioning theory, e.g. those of Flowers and Vuji\v ci\'c \cite{Flo63} and of Br\'emond and Valatin \cite{Bre63}. 
More recently, Civitarese, Reboiro and Vogel \cite{Civ97} concluded that  ``an isospin-symmetric Hamiltonian, treated with the generalized Bogolyubov transformation, fails to describe the ground state properties correctly". This transformation naturally diagonalizes the Hamiltonian in the mean field approximation, i.e. upon the replacement of some operators associated with large matrix elements by their expectation value. In our case, the monopole pair operators $P^\dagger$ are replaced by the pairing anomalous density $P^\dagger\rightarrow \langle P^\dagger\rangle$. The existence of \emph{only pair condensates} in the case of $pn$ pairing should thus be carefully considered.

Nevertheless, the standard mean field approach to pairing in $N = Z$ nuclei  has been for many years the generalized Hartree-Fock-Bogoliubov approximation \cite{Goo79}, where all types of Cooper pairs are treated in a unified manner. In the last decade, the particle number and isospin conserving Quartet Condensation Model (QCM) was proposed for the study of isovector pairing and 
quarteting correlations in $N=Z$ nuclei \cite{San12a,danielphd}. It was further developed in 
Refs. \cite{San12,Neg14,San14,San15,Neg17,Neg18} to the 
case of isoscalar pairing and $N>Z$ nuclei. General microscopic quartet models for a 
shell-model basis with an effective Hamiltonian were also recently proposed \cite{Fu13,Fu15,Sam151,Sam152,Sam153,Sam154}. In these approaches, the basic building blocks are not the Cooper pairs anymore, but four-body structures composed of two neutrons and two protons coupled to the isospin $T = 0$ and to the angular momentum $J = 0$, denoted "$\alpha$-like quartets". Interesting connections between the symmetry restricted pair condensate and the quartet descriptions have very recently been uncovered \cite{romero2019}. However, as to this day, ``no symmetry-unrestricted mean-field calculations of $pn$ pairing, based on realistic effective interaction and the isospin conserving formalism have been carried out" \cite{romero2019, perlinska2004}.

 In this work, we take a step along this direction. We construct a solvable BCS-like theory of quartet correlations without assuming any mean field approximation, but only a coherent state ansatz of BCS type involving the isoscalar quartet operator. At variance with the quartet BCS theory of Schuck et al \cite{schuck2016}, solving our simpler model requires no approximation.

\emph{Formalism}. The system of interest consists of a number $N=Z$ of  neutrons and protons moving outside a self-conjugate inert core, and interacting through a charge-independent pairing force. The isovector pairing Hamiltonian is suitable to describe both spherical and deformed nuclei,
\beq
\label{ham}
H=\sum_{i=1}^{N_\text{lev}}\epsilon_i N_{i,0}+\sum_{\tau=0,\pm1}\sum_{i,j=1}^{N_\text{lev}}V_{ij}P^{\dagger}_{i,\tau } P_{ j,\tau}~,
\eeq
where $i,j$ denote the single particle doubly-degenerate states and $\epsilon_i$ refers to the single particle energies; a time conjugated state will be denoted by $\bar{i}$. The $N_{i,0}$ operator counts the total number of particles,
$N_{i,0}=N_{i,1}+N_{i,-1}=\sum_{\tau=\nu,\pi}\left(c^\dagger_{i,\tau}c_{i,\tau}+c^\dagger_{\bar{i},\tau}c_{\bar{i},\tau}\right)$, and the isovector triplet of pair operators is given by $P^\dagger_{i,1}=c^\dagger_{i,\nu}c^\dagger_{\bar{i},\nu},
P^\dagger_{i,-1}=c^\dagger_{i,\pi}c^\dagger_{\bar{i},\pi},
P^\dagger_{i,0}=\dfrac{1}{\sqrt{2}}\left(c^\dagger_{i,\nu}c^\dagger_{\bar{i},\pi}+c^\dagger_{i,\pi}c^\dagger_{\bar{i},\nu}\right)$.

An almost exact solution for the Hamiltonian (\ref{ham}) may be obtained within the QCM. The formulation of this model is based on correlated four-particle structures known as ``quartets". To obtain a quartet, one first defines a set of collective $\pi\pi$, $\nu\nu$ and $\pi\nu$ Cooper pairs $\Gamma^\dagger_\tau(x)\equiv\sum_{i=1}^{N_\text{lev}}x_i P^{\dagger}_{i,\tau}$, 
which depend on a set of mixing amplitudes $x_i$, $i=1,2,...,N_{\text{lev}}$. A collective quartet operator is then constructed by coupling two collective pairs to the total isospin $T=0$, 
\beq
\label{cohq}
Q^\dagger(x)\equiv\left[\Gamma^\dagger\Gamma^\dagger\right]^{T=0}_{S=0}\equiv 2\Gamma^\dagger_1(x)\Gamma^\dagger_{-1}(x)-\big[
\Gamma^\dagger_0(x)\big]^2~.
\eeq 
For $N=Z$ nuclei, the ground state of the Hamiltonian (\ref{ham}) is taken to be a ``condensate'' of such $\alpha$-like quartets, $| \Psi(x) \rangle=\big[Q^{\dagger}(x)\big]^{n_q}|0\rangle$, where $n_q$ is the number of quartets. The concept of a ``condensate'' denotes here the state obtained by acting with the same operator a number of times on the vacuum. It should not be confused with an ideal boson-type condensate. The model is solved by determining numerically the mixing amplitudes $x_i$ upon the minimization of the Hamiltonian expectation value subject to the unit norm constraint, i.e. $
\delta\langle \Psi (x)| H| \Psi(x)\rangle=0$, $\langle  \Psi(x) | \Psi (x)\rangle=1$. 

 The quartet state has, by construction, a well defined particle number and isospin. As such, it may be well suited to precisely describe various nuclear properties that are sensitive to the fluctuations of these quantum numbers. However, the presence of a condensate is not guaranteed in a particle number conserving theory (which always has a nontrivial solution). With this motivation, we look for a BCS-like formulation of quartet correlations which may provide more insight into the quartet condensation phenomenon. 
 
 In analogy with the standard BCS case, we consider a coherent state involving the quartet operator
 \beq
 \label{QBCSoriginal}
|QBCS\rangle\equiv\exp(Q^\dagger)|0\rangle =\sum_{n=0}^{N_\text{lev}} \frac{1}{n!}(Q^\dagger)^n|0\rangle~.
 \eeq
Despite of the formal similarity with the standard BCS theory, this ansatz is considerably more complicated. To see this, consider a simple BCS state for like-particles, e.g. neutrons 
\beq 
\begin{aligned}
|BCS\rangle&\equiv\exp (\Gamma^\dagger_1)|0\rangle=\exp(\sum_{i=1}^{N_\text{lev}} x_i P^\dagger_{i,1})|0\rangle\\
&=\prod_{i=1}^{N_\text{lev}}(1+x_iP^\dagger_{i,1})|0\rangle~.
\end{aligned}
\eeq
The BCS state can be factorized as the exponent contains a single sum over the single particle states. In the quartet case however, the exponent is expanded as $Q^\dagger=\sum_{i,j=1}^{N_\text{lev}}x_i x_j (2P^\dagger_{i,1}P^\dagger_{j,-1}-P^\dagger_{i,0}P^\dagger_{j,0})$. It is a highly entangled mixture of all single particle states which hinders a direct factorization. To develop a tractable BCS-like theory for quartets, we need a linearization procedure for the exponent.

We achieve this by first writing the coherent quartet as a quadratic form. Consider the rotated pair operators defined as $
p_{j,1}^\dagger=({P_{j,1}^\dagger+P_{j,-1}^\dagger})/{\sqrt{2}}$, $p_{j,2}^\dagger=i({P_{j,1}^\dagger-P_{j,-1}^\dagger})/{\sqrt{2}}$ and
$p_{j,3}^\dagger=i P_{j,0}^\dagger$.
The corresponding coherent pairs are defined by $\gamma_a^\dagger=\sum_{j=1}^{N_\text{lev}} x_j p_{j,a}^\dagger$.
The coherent quartet becomes $Q^\dagger=(\gamma_{1}^\dagger)^2+(\gamma_{2}^\dagger)^2+(\gamma_{3}^\dagger)^2$. At this stage, the quartet-BCS state may be factorized as $|QBCS\rangle= \prod_{a=1}^3\exp[(\gamma^\dagger_a)^2]|0\rangle$.

To proceed, we linearize the exponent in each of the three factors by using the following property of Gaussian integration (see standard textbooks, e.g. \cite{Zee})

\beq
\exp(A^2)=\frac{1}{\sqrt{4\pi}}\int_{-\infty}^{+\infty} \text{d}z\, \exp(-z^2/4+zA) ~.
\eeq
and we write the QBCS state as

\beq
|QBCS\rangle=\frac{1}{(4\pi)^{3/2}}\int\text{d}^3z\exp\left(-\vec{z}^{\,2}/4+\vec{z}\cdot\vec{\gamma}^\dagger \right)|0\rangle~,
\eeq
with $\vec{z}=(z_1,z_2,z_3) \in \mathbb{R}^3$ and $\vec{\gamma}=(\gamma_1,\gamma_2,\gamma_3)$. In this form, the exponent is a simple linear combination of single particle operators, $\vec{z}\cdot\vec{\gamma}^\dagger=\sum_{i=1}^{N_\text{lev}}x_i \vec{z}\cdot\vec{p}_{i}^{\,\dagger}$, which may be factorized just as in the BCS case

\beq
\label{QBCSfinal}
\begin{aligned}
|QBCS\rangle=&\frac{1}{(4\pi)^{3/2}}\int\text{d}^3z\exp\left(-\vec{z}^{\,2}/4 \right) \times \\
&\prod_{i=1}^{N_\text{lev}} (1+x_i \vec{z}\cdot \vec{p}^{\,\dagger}_i+x_i^2 \vec{z}^{\,2}q^\dagger_i/2)|0\rangle~,
\end{aligned}
\eeq
where we took into account that at most second order terms are generated in each factor; they involve the quartet operator that fills completely the level $i$, $q_i^\dagger=(p_{i,a}^\dagger)^2=c^\dagger_{i,\pi}c^\dagger_{\bar{i},\pi}c^\dagger_{i,\nu}c^\dagger_{\bar{i},\nu}$.

Let us note that while the QBCS state violates particle number conservation, it still has a good isospin $T=0$. As such, the integral appearing in Eq. (\ref{QBCSfinal}) may be interpreted as an isospin projection operation acting on a generalized Br\'emond-Valatin factorized state $|\psi(z,\hat{z})\rangle=\prod_{i=1}^{N_\text{lev}} (1+x_i z \hat{z}\cdot \vec{p}^{\,\dagger}_i+x_i^2 z^{\,2}q^\dagger_i/2)|0\rangle$ (see Ref. \cite{Bre63}). Here, at variance with the original Br\'emond-Valatin ansatz, the isovector pair may be oriented along any direction $\hat{z}$ in isospin space. Moreover, the Gaussian integration has the crucial role of performing configuration mixing on the  Br\'emond-Valatin factorized ansatz (which by itself misses most of the relevant correlations). As such, the QBCS state is a superposition of all possible factorized states, each having the relative amplitude of finding a pair on the level $i$ proportional to $x_i$, and that of finding a quartet to $x_i^2$. Note that, while the factorized Br\'emond-Valatin may be treated by a quasi-particle transformation, the integration necessary to obtain the QBCS state completely destroys  the quasi-particle picture.
 
As a first step, in the present work we limit ourselves to assessing the consequences of breaking only the particle number conservation; we leave the breaking of isospin conservation to future investigations. We compute the ground state correlations within the QBCS model by minimizing the expectation value  of  $\mathcal{H}(x)=\langle QBCS|H-\lambda N_0 |QBCS\rangle/\langle QBCS|QBCS\rangle$ subject to the particle number constraint $\langle QBCS| N_0 |QBCS\rangle/\langle QBCS|QBCS\rangle=4n_q$, where $n_q$ is the number of quartets, $n_q=(N+Z)/4$. The expressions of the norm and of the various operator averages on the QBCS state as functions of the mixing amplitudes may easily be extracted from Eq. (\ref{QBCSfinal}).  Here, for illustrative purposes we just compare the QBCS and BCS norm functions in the simplest case of $N_\text{lev}=2$; for the QBCS we obtain  $\langle QBCS |QBCS\rangle=225 x_1^4 x_2^4+12 x_1^2 x_2^2+9 x_1^4+ 9x_2^4+1$, while the BCS norm is just $\langle BCS |BCS\rangle=x_1^2x_2^2+x_1^2+x_2^2+1$.
While the BCS norm may be factorized as $\langle BCS |BCS\rangle=(1+x_1^2)(1+x_2^2)$, the QBCS norm does not admit any factorization.

\begin{figure}[t!]

\includegraphics[width=\columnwidth]{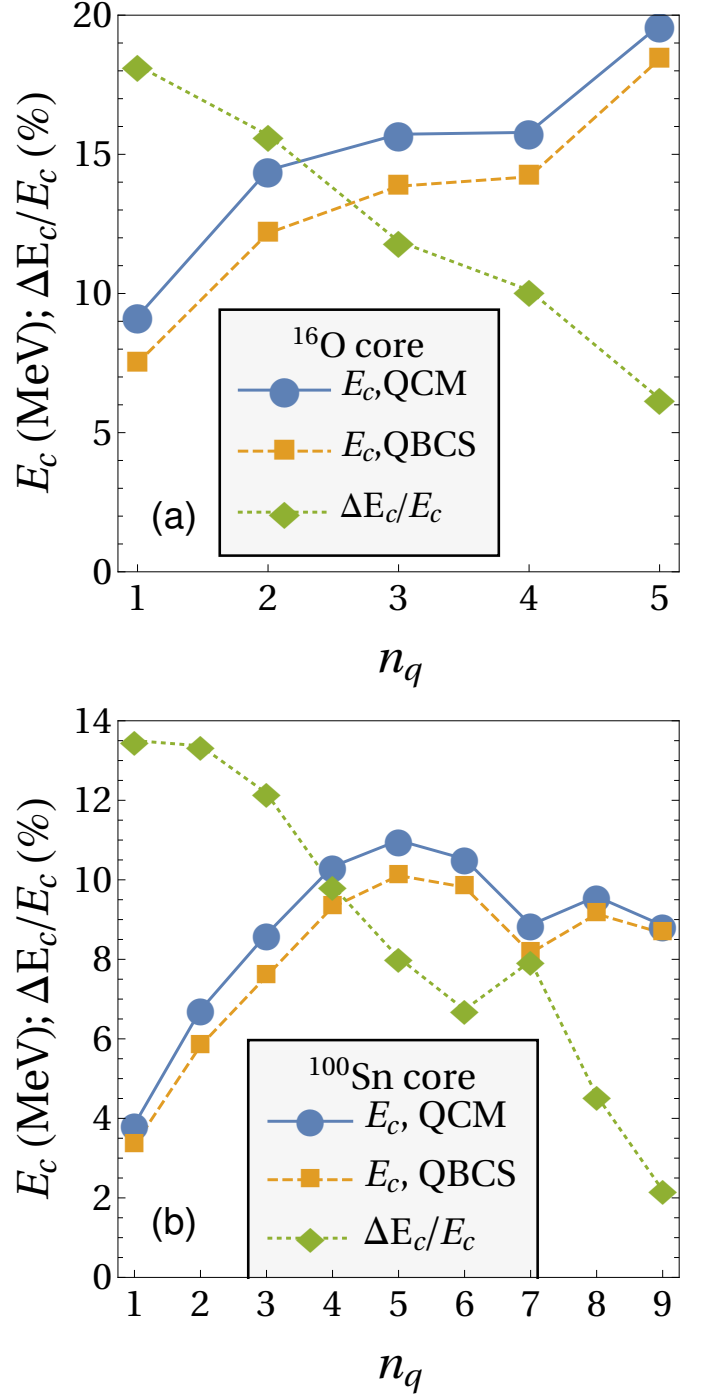}

\caption{Correlation energies $E_{\text{c}}$ in MeV versus the number of quartets $n_q$, for the $^{16}$O core (a) and for the  $^{100}$Sn core (b), together with the relative error $\Delta E_c/E_c=1-E_{c,\text{QBCS}}/E_{c,\text{QCM}}$.  }
\label{fig1}

\end{figure}

 A similar behaviour is valid for the other relevant quantities of the theory, which are given by  more complicated expression in the QBCS case than in the standard BCS (see the general expressions  given in the Supplementary Material).

\begin{figure}[t!]

 \centering

\includegraphics[width=\columnwidth]{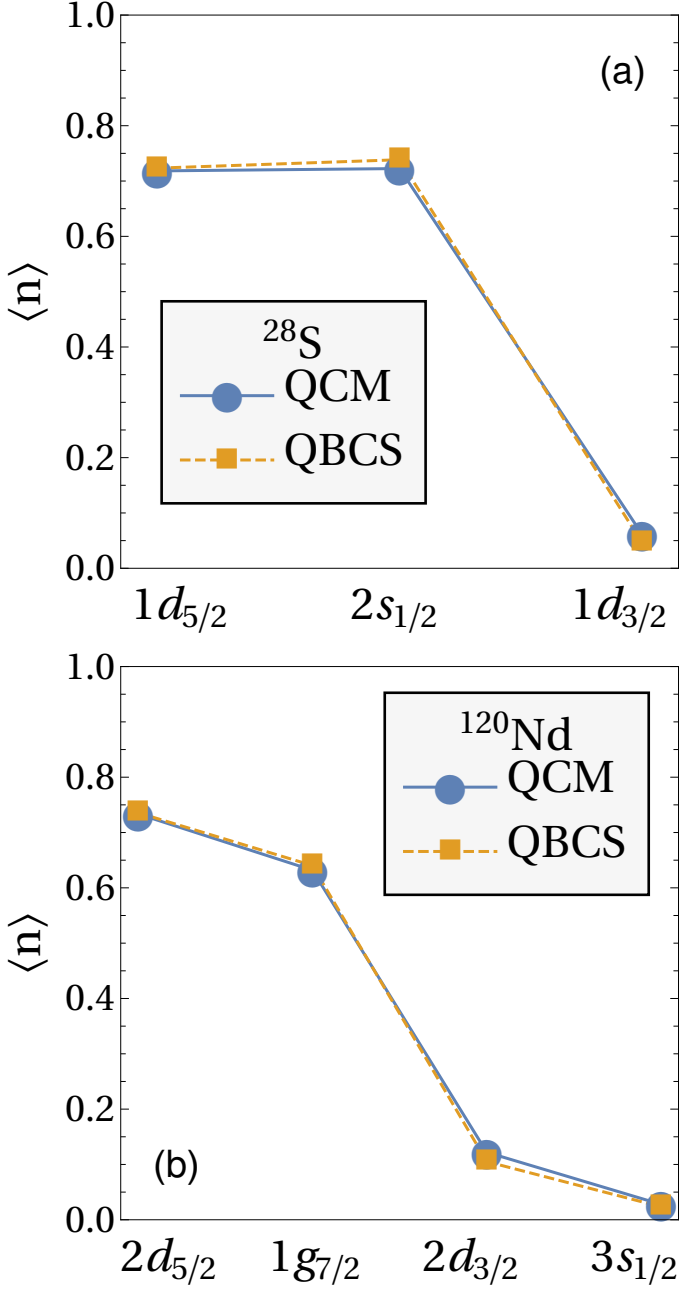}

\caption{Average level occupations $\langle n_{i}\rangle=\langle N_{i,0}\rangle/2(2j_i+1)$, versus the spherical state , for the $^{16}$O core (a) and for the  $^{100}$Sn core (b).}
\label{fig2}

\end{figure}

\emph{Numerical results}. We test our formalism against the well established QCM in the nuclei above $^{100}$Sn and above $^{16}$O. We consider the same model spaces and interactions as in Refs. \cite{San12a,San12}, namely the spherical spectrum $\epsilon_{2d_{5/2}}=0.0$MeV,  $\epsilon_{1g_{7/2}}=0.2$MeV,  $\epsilon_{2d_{3/2}}=1.5$MeV,  $\epsilon_{3s_{1/2}}=2.8$MeV together with the effective Bonn A potential of Ref. \cite{Jen95} for the $sdg$ shell and the spectrum $\epsilon_{1d_{5/2}}=-3.926$MeV, $\epsilon_{2s_{1/2}}=-3.208$MeV, $\epsilon_{1d_{3/2}}=2.112$MeV together with the USDB interaction of Ref. \cite{Bro06} for the $sd$ shell. We solve the QCM model using the analytical method of Refs. \cite{qcmanalitic,qcmanalitic2}.

 We show in Fig. (\ref{fig1}) the results for the correlations energies defined as $E_c = E_0- E$, where $E$ is the energy of the ground state and $E_0$ is the energy calculated without taking into account the isovector pairing interaction. In both cases, we observe  a good agreement between the QBCS and the particle number conserving QCM (note that, at variance with the particle number projected BCS, the QCM offers practically the exact solution, within 1\% error). The relative error $\Delta E_c/E_c$ generally decreases with the increase of the number of quartets, reaching $\mathcal{O}(1\%)$ around $n_q=10$. The only exception is the case with $n_q=7$ quartets above $^{100}$Sn, where the closure of the $2d_{5/2}1g_{7/2}$ subshell leads to a local minimum in the correlation energy and a local maximum in the energy relative error.
 
\begin{figure}[t!]

 \centering

\includegraphics[width=\columnwidth]{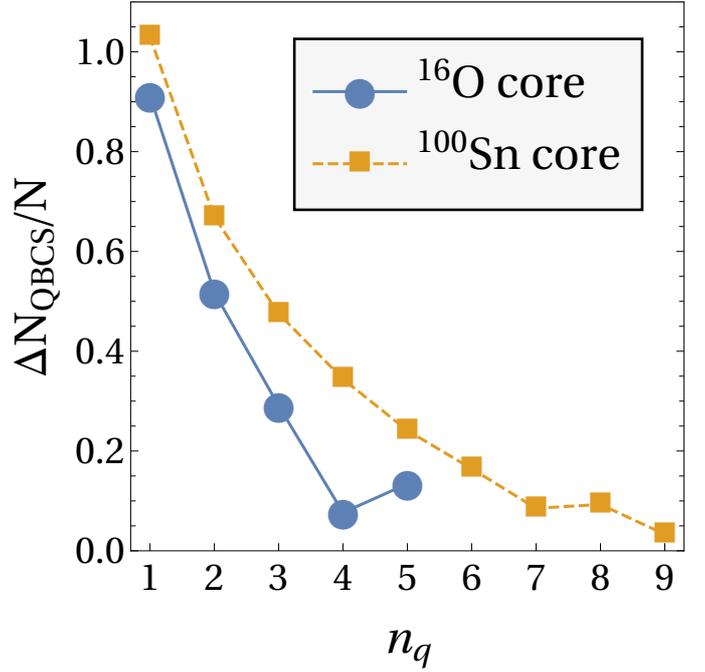}

\caption{ The relative particle number fluctuation versus the number of quartets $n_q$, for the $^{16}$O core and for the  $^{100}$Sn core.}
\label{fig3}

\end{figure}

  A very good agreement is obtained also for the average level occupancies, shown in Fig. (\ref{fig2}) for the nuclei at shell half-filling.
 In Fig. (\ref{fig3}), the relative particle number fluctuation $\Delta N/N=\sqrt{\langle N^2\rangle- N^2}/N$ is shown to decrease with the increase of the number of quartets, much more rapidly than the $1/\sqrt{N}$ rate specific to standard pairing: in both regions the best fit indicates that $\Delta N/N\sim 1/N$.

Although formally very similar, the QCM description of the isovector pairing correlations is much more precise than the particle number projected BCS description for the standard pairing case. As seen above, this remains true also for the coherent state model, where the energy errors are smaller and the particle number fluctuations decrease faster for the QBCS than for the standard BCS. 

This is naturally related to the fact that the QBCS coherent state has components with the particle number a multiple of four, as opposed to two for the standard BCS. We thus expect the component of the coherent state having the correct number of particles to dominate in the quarteting case more strongly than in the standard pairing case, behaviour to be investigated quantitatively in future works.

\begin{figure}[t!]

 \centering

\includegraphics[width=\columnwidth]{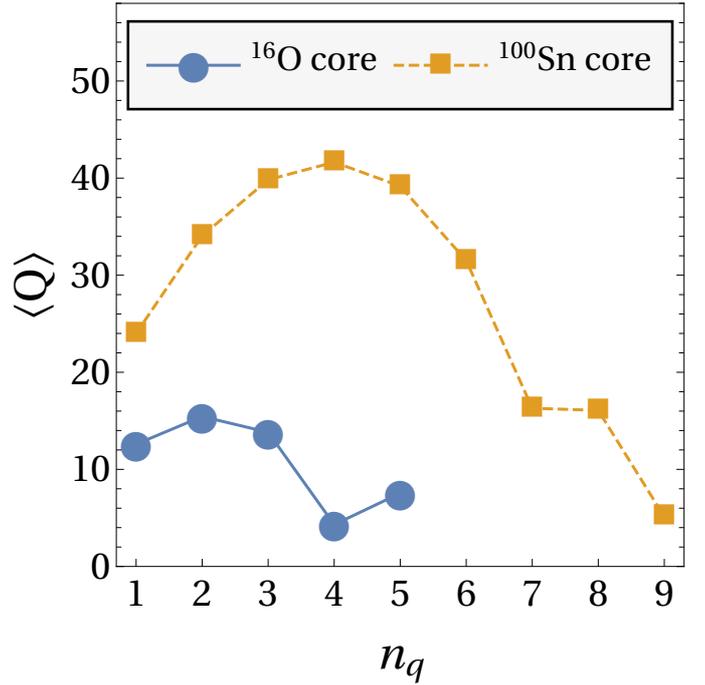}

\caption{ The quartet anomalous density versus the number of quartets $n_q$, for the $^{16}$O core and for the  $^{100}$Sn core.}
\label{fig4}

\end{figure}

 Another consequence of the particular structure of the QBCS state is the absence of the anomalous pair density in our theory, $\langle P\rangle=0$. As an indicator of the four body  correlations, we choose to study the quartet anomalous density
 \beq
 \langle Q\rangle=\langle QBCS|Q(x_i=1)|QBCS\rangle~,
 \eeq
  defined in analogy to the quantity appearing in the standard pairing case, $\langle BCS|\Gamma_{1}(x_i=1)|BCS\rangle$,  related to the pairing gap. Similarly to the standard pairing case, the quartet anomalous density exhibits minima at the open and closed shell configurations and a maximum around shell half-filling, as seen from Fig. (\ref{fig4}). 
  
  Thus far, the pairing and quarteting correlation treated within the BCS and, respectively, QBCS formalisms appear to show quite similar manifestations. The main difference arises when considering the response of the quartet correlations to a varying interaction strength. It is well known that the standard BCS has a nontrivial solution only for interaction strengths greater than a critical value dependent on the model space.

  \begin{figure}[t!]

 \centering

\includegraphics[width=\columnwidth]{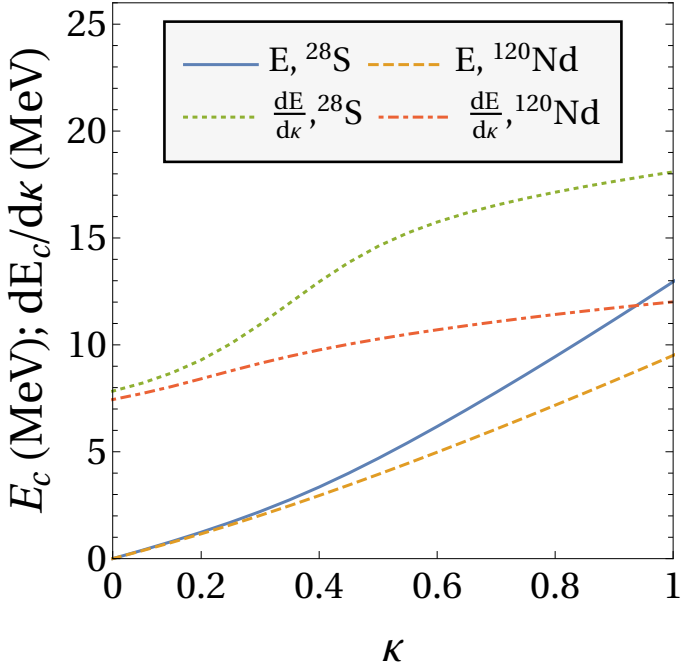}

\caption{The correlation energy and its derivative versus the interaction strenght scaling factor $\kappa$, for the nuclei $^{28}$Si and $^{120}$Nd.}
\label{fig5}

\end{figure}

  In the QBCS case we observe no evidence of a sharp transition from a quarteting to a normal phase as we decrease the interaction strength. As seen from Fig. \ref{fig5}, when scaling the interaction matrix elements by a factor $\kappa\in [0,1]$ the correlation energy varies smoothly, with no jumps of its derivative. Instead, there is a relatively narrow interval towards small values of the scaling factor $\kappa$ where the derivative experiences a more pronounced variation. This behaviour of the QBCS (which does not conserve the particle number) is exhibited also by the number projected BCS theory.  We conclude that the restoration of the isospin symmetry is enough to smoothen the transition from the quarteting to the normal phase.

\emph{Conclusions}. We proposed a BCS-like theory for pure quartet correlations based on a quartet coherent state analogous to the famous $|BCS\rangle$ state. Our ansatz is unique as it does not contain any pairing anomalous density $\langle P \rangle$ (as opposed to usual mean field treatments for $pn$ pairing), but instead one of quartet type $\langle Q\rangle$. We evidenced that the standard pairing and quarteting correlations share similar qualitative features, but there is no sharp quarteting transition. Additionally, we uncovered new connections between the quartet models and some of the early BCS-like attempts to $pn$ pairing.

 The proposed QBCS theory is flexible and there are numerous ways in which it may be easily expanded. Firstly, in order to incorporate the interplay of pairing and quarteting effects, a generalized ansatz exhibiting also a pair condenstate may be considered in the form

  \beq
  |(P\&Q)BCS\rangle=\exp[\hat n\cdot \vec{\Gamma}^\dagger(y)+Q^\dagger(x)]|0\rangle,
  \eeq
  involving an additional coherent isovector pair oriented along a general direction $\hat{n}$ in isospin space. On the other hand, the pure quartet correlations induced by the combined isovector and isoscalar interactions may be treated with the ansatz
    \beq
  |QBCS(iv+is)\rangle=\exp(Q_{iv}^\dagger(x)+Q_{is}^\dagger(y))|0\rangle,
  \eeq
  where $Q_{iv}$ is the quartet considered in this work (built from two isovector pairs) while $Q_{is}$ is the quartet operator built from two isoscalar pairs.
  
  The excited states of the isovector pairing hamiltonian within the QBCS framweork may be computed in several ways, to be analyzed in detail in future works. Although a quasiparticle picture would be preferred, recent results also indicate that it may be inappropriate for quartet correlations \cite{Sogo2010}. Nevertheless, an approximate quasiparticle description may be obtained within a boson formalism along the lines of Ref. \cite{delrjp}. Note that, in deriving the expressions for the averages of the various operators on the QBCS states starting from Eq. (\ref{QBCSfinal}), one may as well treat  the pair operators $p_i$ as bosons and still obtain the exact fermionic results.

Let us note that, just like the BCS equations are simpler than those of the particle number projected BCS, the QBCS equations present a reduced degree of complexity relative to the QCM ones. While we had to use the simpler hole picture (see Ref. \cite{qcmanalitic}) to solve the QCM equations for a large number of quartets ($n_q=7,8,9)$, the QBCS solution was obtained direcly in all cases. As such, the QBCS approach may be used as a simplified approximate framework for the description of quartet correlation; also, it could be used to stabilize the complex nonlinear minimization procedure of the particle number projected QCM, especially for a significant number of quartets.

  Lastly, let us remark that the coherent quartet ansatz of Eq. (\ref{cohq}) is still rather restrictive. More generally, one should consider the case of non-separable mixing coefficients, i.e. $\mathcal{Q}^\dagger=\sum_{i,j=1}^{N_\text{lev}}X_{ij} (2P^\dagger_{i,1}P^\dagger_{j,-1}-P^\dagger_{i,0}P^\dagger_{j,0})$ with $X_{ij}\neq x_i x_j$.This more complex case may also be treated by  taking advantage of the properties of Gaussian integration. In this case we obtain
  \beq
  \begin{aligned}
|QBCS\rangle=&\exp(\mathcal{Q}^\dagger)|0\rangle=\exp(\sum_{i,j}\vec{p}_i X_{ij}\vec{p}_j)  |0\rangle\\
=&\int \text{d}^{3N_{\text{lev}}}z~\exp\left[-\sum_{i,j}\vec{z}_i (X^{-1})_{ij} \vec{z}_j\right] ~\times \\
&
\prod_{i=1}^{N_\text{lev}} (1+\vec{z}_i\cdot \vec{p}^{\,\dagger}_i+\vec{z}^{\, 2} q^\dagger_i/2)|0\rangle~,
\end{aligned}
  \eeq
where a scalar product is understood in all expressions involving two vectors.  The norm and the averages of the various operators on the QBCS state may be automatically computed as Wick contractions of various $X_{ij}$'s by using the Wick theorem for multidimensional Gaussian integrals \cite{Zee}.

\emph{Acknowledgements}. We thank F. Spineanu, M. Vlad,  P. Schuck and J. Dukelsky for valuable discussions and observations. This work was supported by the grants of the Romanian Ministry of Research and Innovation, CNCS - UEFISCDI, PN-
III-P4-ID-PCE-2016-0092, PN-III-P4-ID-PCE-2016-0792, within PNCDI III, and PN-19060101/2019.



  \bibliographystyle{elsarticle-num} 
  \bibliography{mybibfile}


\newpage
\section*{Supplementary Material}
We present here some computational details regarding the general expressions of the norm and various operator averages in the QBCS theory.\\

The norm of the QBCS state may be written as a double integral
\beq
\begin{aligned}
&\langle QBCS |QBCS\rangle=\frac{1}{(4\pi)^{3}}\int \text{d}^3w~\text{d}^3z\exp\left(-\vec{z}^{\,2}/4 -\vec{w}^{\,2}/4 \right)\times\\
&  \prod_{i=1}^{N_\text{lev}}\langle 0 [1+x_i \vec{w}\cdot \vec{p}_i+x_i^2 \vec{w}^{\,2} q^\dagger_i/2]  [1+x_i \vec{z}\cdot \vec{p}^\dagger_i+x_i^2 \vec{z}^{\,2} q^\dagger_i/2]|0\rangle\\
&=\frac{1}{(4\pi)^{3}}\int \text{d}^3w~\text{d}^3z\exp\left(-\vec{z}^{\,2}/4 -\vec{w}^{\,2}/4 \right)\times\\
&  \prod_{i=1}^{N_\text{lev}} (1+x_i^2\sum_{a,b=1}^3 w_a z_b \langle0| p_{i,a} p^\dagger_{i,b}|0\rangle +x_i^4 \vec{w}^{\,2} \vec{z}^{\,2} \langle 0| q_{i} q^\dagger_{i}|0\rangle/4 ) \\
&=\frac{1}{(4\pi)^{3}}\int \text{d}^3w~\text{d}^3z\exp\left(-\vec{z}^{\,2}/4 -\vec{w}^{\,2}/4 \right)\times\\
& \prod_{i=1}^{N_\text{lev}} (1+x_i^2 \vec{w}\cdot\vec{z} +x_i^4 \vec{w}^{\,2} \vec{z}^{\,2}/4 ) ~.
\end{aligned}
\eeq

The integration may be trivialized by passing to spherical coordinates,
\beq
\begin{aligned}
&\langle QBCS |QBCS\rangle\\
&=\frac{1}{8\pi}\int_0^\infty \text{d}w~ w^2 \exp\left(-{w}^{2}/4 \right)  \int_0^\infty\text{d}z ~z^2 ~\exp\left(-{z}^{2}/4 \right)\times \\
&\int_{-1}^1\text{d}(\cos \theta) \prod_{i=1}^{N_\text{lev}} (1+x_i^2 wz\cos\theta +x_i^4 {w}^{2} {z}^{2}/4 ) ~.
\end{aligned}
\eeq

Using the well known properties of Gaussian integration, this expression (and those presented below) may be easily evaluated for any number of levels. The single particle term reads

\beq
\begin{aligned}
&\langle QBCS |N_{0,k}|QBCS\rangle\\
&=\frac{1}{(4\pi)^{3}}\int \text{d}^3w~\text{d}^3z\exp\left(-\vec{z}^{\,2}/4 -\vec{w}^{\,2}/4 \right)\times \\
&( 2 x_k^2 \vec{w}\cdot\vec{z} +x_k^4 \vec{w}^{\,2} \vec{z}^{\,2} )
\prod_{i\neq k} (1+ x_i^2 \vec{w}\cdot\vec{z} +x_i^4 \vec{w}^{\,2} \vec{z}^{\,2}/4 ) ~,
\end{aligned}
\eeq

The interaction term is decomposed into diagonal and nondiagonal parts,
\beq
H_{\text{int}}=\sum_{i,j=1}^{N_\text{lev}}V_{ij}\sum_{a=1}^3 p^{\dagger}_{i,a}p_{j,a}=\sum_{i=1}^{N_\text{lev}}V_{ii}\sum_{a=1}^3 p^{\dagger}_{i,a}p_{i,a}+\sum_{i\neq j}V_{ij}\sum_{a=1}^3 p^{\dagger}_{i,a}p_{j,a}~.
\eeq

The diagonal part is
\beq
\begin{aligned}
&\langle QBCS |\sum_{k}V_{kk}\sum_{a=1}^3 p^{\dagger}_{k,a}p_{k,a} |QBCS\rangle\\
&=\frac{1}{(4\pi)^{3}}\int \text{d}^3w\text{d}^3z\exp\left(-\vec{z}^{\,2}/4 -\vec{w}^{\,2}/4 \right)\times\\
& \sum_{k=1}^{N_\text{lev}} V_{kk} (x_k^2 \vec{w}\cdot\vec{z} +3x_k^4 \vec{w}^{\,2} \vec{z}^{\,2}/4 ) \prod_{i\neq k} (1+ x_i^2 \vec{w}\cdot\vec{z} +x_i^4 \vec{w}^{\,2} \vec{z}^{\,2}/4 ) 
\end{aligned}
\eeq

The non-diagonal part is
\beq
\begin{aligned}
&\langle QBCS |\sum_{k\neq l}V_{kl}\sum_{a=1}^3 p^{\dagger}_{k,a}p_{l,a} |QBCS\rangle\\
&=\frac{1}{(4\pi)^{3}}\int \text{d}^3w\text{d}^3z\exp\left(-\vec{z}^{\,2}/4 -\vec{w}^{\,2}/4 \right)\times\\
&  \sum_{k\neq l} V_{kl} [x_kx_l (1+x_k^2x_l^2w^2z^2/4)\vec{w}\cdot\vec{z}+x_kx_l(x_k^2+x_l^2)z^2w^2/2]\times\\
& \prod_{i\neq k,l} (1+ x_i^2 \vec{w}\cdot\vec{z} +x_i^4 \vec{w}^{\,2} \vec{z}^{\,2}/4 ) 
\end{aligned}
\eeq

Similar expressions may be easily obtained  for the quartet anomalous density and for the square of the total particle number operator.

\end{document}